\documentclass[twocolumn,showpacs,preprintnumbers,amsmath,amssymb]{revtex4}

\usepackage{graphicx}
\usepackage{dcolumn}
\usepackage{bm}
\usepackage{amsmath}
\usepackage{amssymb}

\begin{document}

\preprint{}

\title{ On supersolid fraction}
 
\author{Yongle Yu}
 \affiliation{State Key Laboratory of Magnetic Resonance
  and Atomic and Molecular Physics, Wuhan Institute of Physics and Mathematics,
 CAS,  Wuhan 430071, P. R. China}


\begin{abstract}

We present an  explanation of
 why the observed supersolid fractions of 
helium solids are rather far below unity.   
One might observe 
large supersolid fraction of
neon systems immersing in liquid $^3$He. A system
of bosonic ions in a ring trap could 
display a supersolid fraction close to unity.

\end{abstract}

\pacs{67.80.-s}
\maketitle   

\begin{figure}[htp]

\includegraphics
{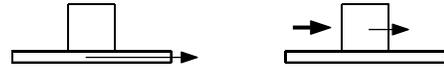}
\caption{ A supersolid crystal (square) is frictionless (left).
But it moves under a normal force (right).
}
\label{normal_force}
\end{figure}

A supersolidic phase of matter is 
characterized by nonclassical
rotational inertia (NCRI) of solid $^4$He 
\cite{chan, kubotaetal,  reppy2}. The observed
 NCRI fractions
are rather far below unity and 
  show  dependence
  on the  crystal perfection. The largest NCRI 
  fraction is 
around 20 percents with a rapid freezing solid
 $^4$He  \cite{reppy2}. These seem to suggest that
supersolidity is related to disorder in solid. 
For the following
considerations on superfluidity and on crystals, 
however, 
  the observation
of rather small
NCRI fractions could be explained and the
 unclear disorder-related
mechanism of supersolidity might be unnecessary. 
 Superfluidity means
that a system can decouple from the motion of
its environment. A superfluidic
crystal is frictionless  given that 
temperature and velocity are low enough.
 However, having a rather fixed shape 
 and regular facets, a superfluidic
 crystal responds to
 a normal force and  moves  (see Fig. \ref{normal_force}). 
 In typical supersolid 
 experiments with  solid $^4$He
  confined in an annual region, the walls
  contacting the solid are made cylindrical. 
  However, the walls, which are solid themselves,
  microscopically can't have a perfect cylindrical
  surface.  Therefore,
  when the walls rotate, the spatial place (position) 
  taken by the walls don't remain the same, and the walls
   exert some normal force
   on part of solid $^4$He.
    The part of 
   the solid which moves together
  with walls behaves like normal solid, 
   although  it might be supersolid. 
   
   The above considerations could 
   also explain that the observed supersolid 
   fraction decreases with annealing \cite{reppy2}.
   The smaller grains of crystal in
  the solid are, the less portion of the solid 
  faces normal force brought by the motion of
  the walls.  
  
  The situation is different in the
  case of liquid $^4$He confined within an annular region.
  Only a very thin layer of liquid
  close to the walls might face the normal force
  brought by the motion of the walls, while the main
  part of liquid  can  remain static.

  We recently discuss the possibility that solid $^{20}$Ne or
 solid $^{22}$Ne, having similar quantum character to that of
  solid $^4$He, could
  possess a supersolid phase \cite{yu}. 
 It might be possible that
 one changes the experimental setup a little bit and
 observes rather large supersolid fraction in neon systems. One could
  fill an annual region 
  with a mixture of liquid $^3$He  and small grains of
  neon crystals.
  The neon crystals could then mostly response to the motion of
  liquid $^3$He and largely decouple from the motion of 
  the walls.
  The liquid $^3$He shall be in the normal phase.
  the transition temperature of liquid $^3$He 
  under ambient pressure is around 1 mK, which is
  likely lower than transition temperature
  of possible supersolidic neon.
  
  Supersolidity could be also realized in another system: a
  number of identical bosonic ions confined in a ring trap.
  The crystalline phase of ions in a trap has long 
  being observed \cite{ions}. When identical Bose 
  ions rotate in 
  a ring, the system could demonstrate a frictionless motion
  with a supersolid fraction close to unity near
  zero temperature. 
   

 \end{document}